\documentclass[manuscript=fullresearchpaper]{beilstein}

\usepackage{graphicx}
\usepackage{multirow}
\usepackage{tabularx}

\begin{document}
\title{In-situ growth optimization in focused electron-beam induced deposition}

\author[]{Paul M. Weirich}
\author[]{Marcel Winhold}
\author*{Christian H. Schwalb}{schwalb@em.uni-frankfurt.de}
\author[]{Michael Huth}
\affiliation{Physikalisches Institut, Goethe Universit\"at, Max-von-Laue-Str. 1, 60438 Frankfurt am Main, Germany}

\maketitle

\begin{abstract}
We present the application of an evolutionary genetic algorithm for the \textit{in-situ} optimization of nanostructures prepared by focused electron-beam-induced deposition. It allows us to tune the properties of the deposits towards highest conductivity by using the time gradient of the measured \textit{in-situ} rate of change of conductance as fitness parameter for the algorithm. The effectiveness of the procedure is presented for the precursor $\rm W(CO)_6$ as well as for post-treatment of Pt-C deposits obtained by dissociation of $\rm MeCpPt(Me)_3$. For $\rm W(CO)_6$-based structures an increase of conductivity by one order of magnitude can be achieved, whereas the effect for $\rm MeCpPt(Me)_3$ is largely suppressed. The presented technique can be applied to all beam-induced deposition processes and has great potential for further optimization or tuning of parameters for nanostrucures prepared by FEBID or related techniques. 
\end{abstract}


%%%%%%%%%%%%%%%%%%%%%%%%%%%%%%%%%%%%%%%%%%%%%%%%%%%%%%%%%%%%%%%
% Introduction
%%%%%%%%%%%%%%%%%%%%%%%%%%%%%%%%%%%%%%%%%%%%%%%%%%%%%%%%%%%%%%

\section{Introduction}

In focused electron-beam-induced deposition, FEBID in short, a (metal-) organic or inorganic volatile precursor gas, previously adsorbed on a substrate surface, is dissociated in the focus of an electron beam provided by a scanning (SEM) or transmission electron microscope (TEM). During the last decade FEBID has developed from a highly specialized nanofabrication method with a limited selection of application fields to one of the most flexible approaches for functional nanostructure fabrication with true 3D patterning capabilities. By now FEBID-based nanostructures are used in highly miniaturized magnetic field \cite{Boero2005,Serrano-Ramon2011}, strain/force \cite{Schwalb2010a, Huth2012} and gas sensing \cite{Kolb2013} applications, as well as in micromagnetic studies on domain wall nucleation and propagation \cite{Fernandez-Pacheco2009,Serrano-Ramon2013}. On the basis of the in-situ, electron irradiation-induced tunability of metallic FEBID- structures significant progress could be made in understanding the charge transport regimes in nanogranular metals \cite{Sachser2011,Porrati2011,Porrati2009}. In addition, by the controlled combination of two precursors it has become possible to prepare amorphous binary alloys \cite{Winhold2011,Porrati2012}, as well as nanogranular intermetallic compounds \cite{Che2005}.

As the FEBID-immanent parameter space becomes larger, the identification of the parameters for an optimized deposition protocol is becoming a very challenging problem. In fact, even for a single organometallic precursor, finding the deposition parameters for, e.g., obtaining the maximum possible metal content, can be a difficult task. This can be exemplified for the commonly used precursor $\rm W(CO)_6$. Rosenberg and co-workers recently studied the electron-dose and substrate-temperature dependence on the final deposit in electron-induced dissociation experiments with 500 eV electron energy for this precursor \cite{Rosenberg2012,Rosenberg2013}. They showed that the initial dissociation at electron doses below about 100 $\rm pC/\mu m^2$ leads to the release (i.e., dissociation and desorption) of two CO ligands from the parent molecule. The decarbonylated residual species is then subject to electron-stimulated decomposition rather than desorption resulting in an average composition of the deposit of [W]/[C]  $\sim$ 1/4. By increasing the electron dose and/or the substrate temperature, which causes changes in the coverage and average residence time of the precursor molecules on the surface, the metal content can be increased to above 40 at\% \cite{Mulders2011}. Changes of the precursor flux and the partial pressure of water in the residual gas also influence the final composition and increases the extend of tungsten oxidation in the deposit \cite{Rosenberg2013}. 

With regard to the electrical conductivity of the deposits, a key quantity in many applications of FEBID structures, no reliable prediction can be made concerning its value for different deposition parameters and conditions. This is due to the fact that metal content alone is not a sufficient indicator since in most instances transport is of the hopping type, so that the matrix composition and the oxidation state of the metal are also important but a-priori unknown quantities \cite{Huth2012, Sachser2011}. From this one can conclude that the optimization of any FEBID process towards the largest possible conductivity should ideally monitor the conductance as the growth proceeds \cite{Porrati2009} and use this information in adaptively changing the deposition parameters. Here, we present a first implementation of such a feedback control mechanism and employ an evolutionary genetic algorithm (GA) for the in-situ optimization of the electrical conductivity of nanostructures prepared by FEBID \cite{Patent}. By using the time gradient of the measured in-situ conductance as a fitness parameter for the GA we are able to tune the properties of the deposits towards highest conductivity. In order to demonstrate the efficiency of this method, we chose $\rm W(CO)_6$. Our study reveals that an increase of conductivity by two orders of magnitude can be achieved with the GA by solely varying the process parameters pitch p and dwell-time $t_D$ in the deposition process. The precursor-specific limitations of the approach are also exemplified for another precursor, $\rm MeCpPt(Me)_3$, which is known to show only one bond-cleavage in the initial step \cite{Landheer2011}. This results in a largely deposition parameter independent Pt/C ratio. Furthermore, in contrast to tungsten, platinum is not susceptible to oxidation or carbide formation, which results in a nano-granular rather than amorphous microstructure.

%%%%%%%%%%%%%%%%%%%%%%%%%%%%%%%%%%%%%%%%%%%%%%%%%%%%%%%%%%%%%%%
% Experimental
%%%%%%%%%%%%%%%%%%%%%%%%%%%%%%%%%%%%%%%%%%%%%%%%%%%%%%%%%%%%%%%

\section{Experimental}

The FEBID process takes place in a dual-beam SEM/FIB microscope (FEI, Nova Nanolab 600) equipped with a Schottky electron emitter. The precursor gases are introduced into the high-vacuum chamber via a gas injection system through a thin capillary (\O~=~0.5~mm) in close proximity to the focus of the electron beam. As a substrate material n-doped Si(100) (350~$\mu$m)/LPCVD Si$_3$N$_4$ (300~nm) was used equipped with 10/200 nm thick Cr/Au contacts with a separation of 3 $\mu$m that were prepared using UV-lithography and a lift-off process. 

\begin{figure}
	%\centering
	\dblcolfigure{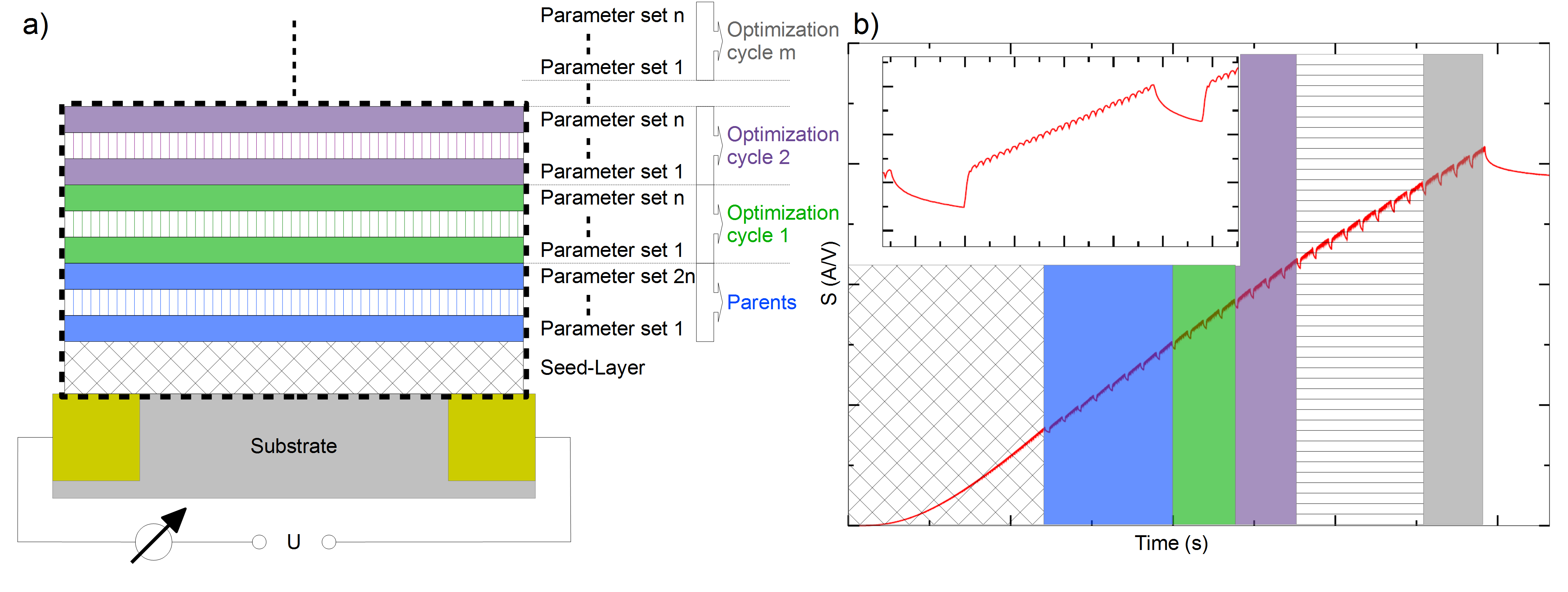}
	\caption{Schematic representation of the optimization process: a) Layer structure of FEBID deposits: m optimization cycles, each consisting of n parameter sets except for the parent optimization cycle with 2n parameter sets, are deposited onto a seed-layer between two Cr/Au electrodes. During the deposition process the conductance of the whole layer structure is measured. b) Representative $S(t)$-graph for layer structure of a). Altering background colors indicate the deposition of different optimization cycles. The inset depicts $S(t)$ during the deposition of one layer. The $S(t)$-curve shows a sharp increase when the FEBID process is started and decreases when the deposition process is stopped, respectively.}
	\label{fig:Figure1}
\end{figure}

The optimization process using the GA in combination with \textit{in-situ} electrical conductance measurements is schematically displayed in Figure \ref{fig:Figure1}a. At first a seed-layer is deposited ensuring that all optimization processes start with the same initial conditions. On top of the seed layer subsequent layers with different deposition parameters are added. 

%%%%%%%%%%%%%%%%%%%%%%%%%%%%%%%%%%%%%%%%%%%%%%%%%%%%%%%%%%%%%%%
% Figure 1
%%%%%%%%%%%%%%%%%%%%%%%%%%%%%%%%%%%%%%%%%%%%%%%%%%%%%%%%%%%%%%%
%%%%%%%%%%%%%%%%%%%%%%%%%%%%%%%%%%%%%%%%%%%%%%%%%%%%%%%%%%%%%%%

\begin{figure}
	%\centering
	\dblcolfigure{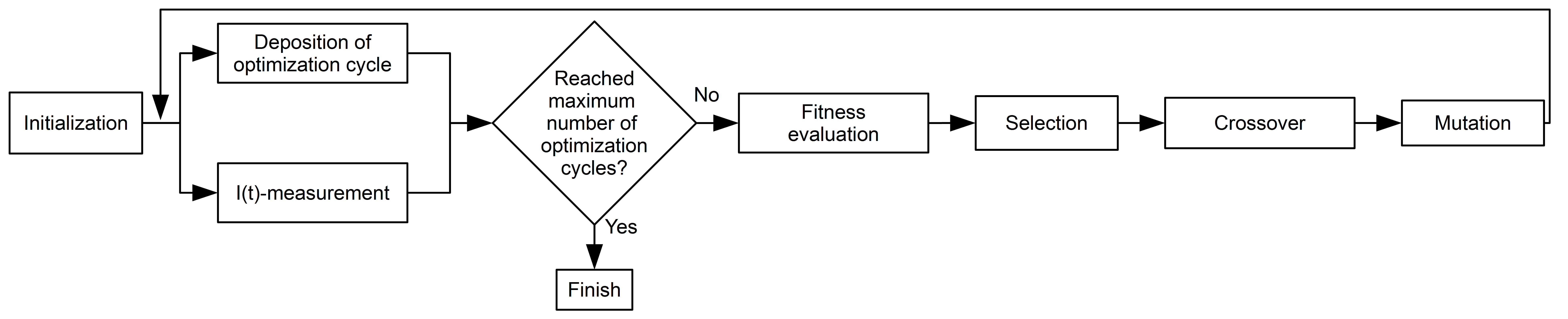}
	\caption{Logical flow representation for the \textit{in-situ} optimization of conductance of FEBID deposits with a GA. After the initialization of the program, the GA optimizes the conductance of the deposits by using the measured gradient of $\overline{S}'$ to evaluate the fitness of the parameter sets used for deposition. Selection, recombination and mutation of parameter sets are carried out after the fitness evaluation to obtain the next optimization cycle with optimized parameter sets. The process is stopped after the deposition of a pre-defined number of optimization cycles.}
	\label{fig:Figure2}
\end{figure}

With regard to a GA-based optimization process, the set of parameters used for the deposition of one specific layer consists of \{x- and y-size of the deposit, dwell time ($t_D$), pitch in x ($p_x$) and y ($p_y$) direction, beam current ($I$), acceleration voltage ($U$), temperature ($T$), refresh-time ($t_r$), scan-type (raster or serpentine), dose ($D$) and passes ($p$)\}. However, not all parameters are independent, e.g. in order to keep $D$ fixed, $P$ has to be adapted according to the specific combination of \{x- and y-size of the deposit, $p_x$ and $p_y$, $I$ and $t_D$ \}. The aim of the GA's search is to find parameter sets leading to an enhancement of conductance due to an increasing growth rate of the deposit and/or intrinsic effects e.g. the increase of the metal content and/or a change of the dielectric matrix. The GA allows for the optimization of deposition parameters for an arbitrary precursor, without having any additional information about the deposition process. Therefore, the following procedure is performed: 

The parent optimization cycle based on the first 2n parameter sets with randomly generated parameters is deposited onto the seed layer. After the deposition of each layer a fitness evaluation is carried out for each parameter set according to the following principle. During the optimization process the conductance S is measured and the rate of change of conductance over time $\overline{S}'$ = $S/t$ is calculated. Assuming a parallel circuited resistance is added, once another layer is deposited on top of the existing structure, $\overline{S}'$ is constant if the growth rate and the conductivity do not change. However, if either the conductivity or the growth rate is altered by the variation of deposition parameters, the gradient of $\overline{S}'$ is a suitable variable to describe the influence of deposition parameters on the conductance of the deposit. Hence, the gradient of $\overline{S}'$ is chosen as the fitness parameter for the GA in order to detect effects leading to a change of the growth rate and/or the conductivity. Layers with the highest fitness values are selected to generate the next optimization cycle of n parameter sets using genetic operators such as crossover and mutation. For the next optimization cycle a number of new parameter sets are created, according to half the size of the initial parent optimization cycle. One half of the next optimization cycle is created with the crossover method, the other half with the mutation method. The parents of the new parameter sets are chosen via an uniform distributed random choice. The crossover method is performed by exchanging parameters of the parents. For the mutation method parameters are chosen randomly within the given parameter-range. A representative time-dependent development of the conductance during the optimization process is shown in Figure \ref{fig:Figure1}b. The GA is stopped after a predefined number of m optimization cycles yielding a set of FEBID deposition parameters for each precursor for a deposit of optimized conductance. A flow-chart of the GA optimization process is shown in Figure \ref{fig:Figure2}.

%%%%%%%%%%%%%%%%%%%%%%%%%%%%%%%%%%%%%%%%%%%%%%%%%%%%%%%%%%%%%%%
% Figure 2
%%%%%%%%%%%%%%%%%%%%%%%%%%%%%%%%%%%%%%%%%%%%%%%%%%%%%%%%%%%%%%%
%%%%%%%%%%%%%%%%%%%%%%%%%%%%%%%%%%%%%%%%%%%%%%%%%%%%%%%%%%%%%%%

%%%%%%%%%%%%%%%%%%%%%%%%%%%%%%%%%%%%%%%%%%%%%%%%%%%%%%%%%%%%%%%
% Results
%%%%%%%%%%%%%%%%%%%%%%%%%%%%%%%%%%%%%%%%%%%%%%%%%%%%%%%%%%%%%%%

\section{Results}

\begin{figure}
	%\centering
	\dblcolfigure{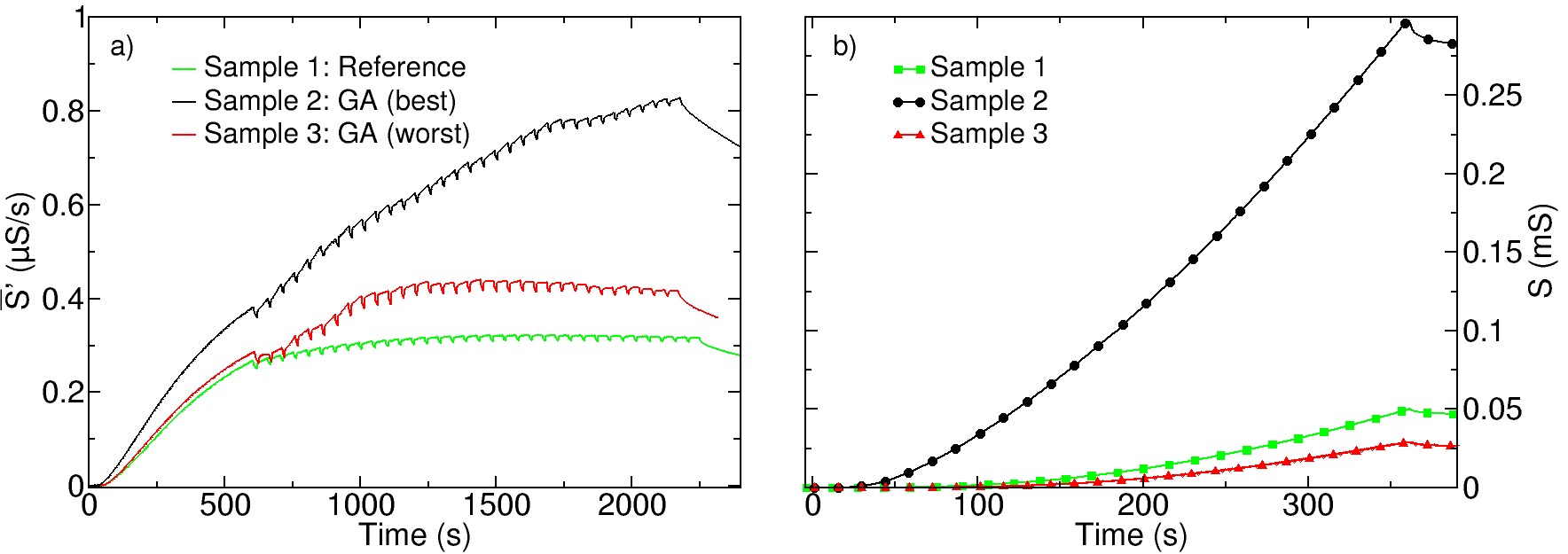}
	\caption{a) Rate of change of conductance during the GA optimization for W-C-O reference (green), GA optimized deposit for highest conductance (black) and GA optimized deposit for lowest conductance (red). For each parameter set a dose of $3~nC/\mu m^2$ was used. The population size amounted to 8 parameter sets and 6 optimization cycles which were deposited for the GA optimization. b) Conductance of $A=3 \times 7~\mu m^2$ W-C-O structures deposited with parameters derived from the optimization processes in a) as well as for the W-C-O reference using a dose of $27~{nC}/{\mu m^2}$.}
	\label{fig:Figure3}
\end{figure}

In order to check for the proper operation of the GA we first applied it for the optimization of deposition parameters for the widely used precursor $\rm W(CO)_6$ \cite{Porrati2009, Hoyle1993, Bauerdick2006, Huth2009}. For $\rm W(CO)_6$ it is well known that the metal content and, respectively, the conductivity strongly depend on the deposition parameters during the FEBID process. At the beginning a reference sample was deposited using standard deposition parameters ($U=5~kV$, $I=6.3~nA$, $t_D=100~\mu s$, $p_x=40~nm$, $p_y=40~nm$). For the reference the GA protocol was used, meaning that the process was paused after the deposition of each layer, indicated by drops in the curves of Figure \ref{fig:Figure3}a. However, for the reference sample the parameters were kept fixed for the complete deposition process. For each parameter set a dose of $3~nC/\mu m^2$ was used. The GA was carried out for 6 optimization cycles with a population size of 8 parameter sets. The measured rate of change of conductance during the FEBID process for the reference sample is displayed in Figure \ref{fig:Figure3}a (Sample 1). Subsequently the GA was applied for finding the optimized parameters for deposition using $\rm W(CO)_6$ as a precursor. First, only the dwell time $t_D$ was used as optimization parameter and was allowed to vary in the range of $0.2-1500~\mu s$. The corresponding rate of change of conductance is displayed in Figure \ref{fig:Figure3}a (Sample 2). In addition, we let the GA search for deposition parameters leading to minimum conductance. Dwell time $t_D$ and pitch $p_x$, $p_y$ were allowed to vary in the range of $0.2-1500~\mu s$ and $30-200~nm$, respectively (Figure \ref{fig:Figure3}a, Sample 3). The highest conductance for W-C-O deposits was obtained for short dwell times ($t_D=0.5~\mu s$) whereas a low conductance was observed for long dwell times ($t_D=831~\mu s$) and a larger y-pitch ($p_y=150~nm$). In order to study the success of the GA procedure the optimized parameter sets returned by the GA for highest and lowest conductance as well as for the reference sample were used for a standard FEBID process and the conductance during deposition was measured (see Figure \ref{fig:Figure3}b). As can be clearly seen, sample 2 (optimized for highest conductance) shows by far the highest value of conductance, whereas for sample 3 (optimized for lowest conductance) the lowest value is achieved.

%%%%%%%%%%%%%%%%%%%%%%%%%%%%%%%%%%%%%%%%%%%%%%%%%%%%%%%%%%%%%%%
% Figure 3
%%%%%%%%%%%%%%%%%%%%%%%%%%%%%%%%%%%%%%%%%%%%%%%%%%%%%%%%%%%%%%%
%%%%%%%%%%%%%%%%%%%%%%%%%%%%%%%%%%%%%%%%%%%%%%%%%%%%%%%%%%%%%%%

\begin{figure}
	%\centering
	\dblcolfigure{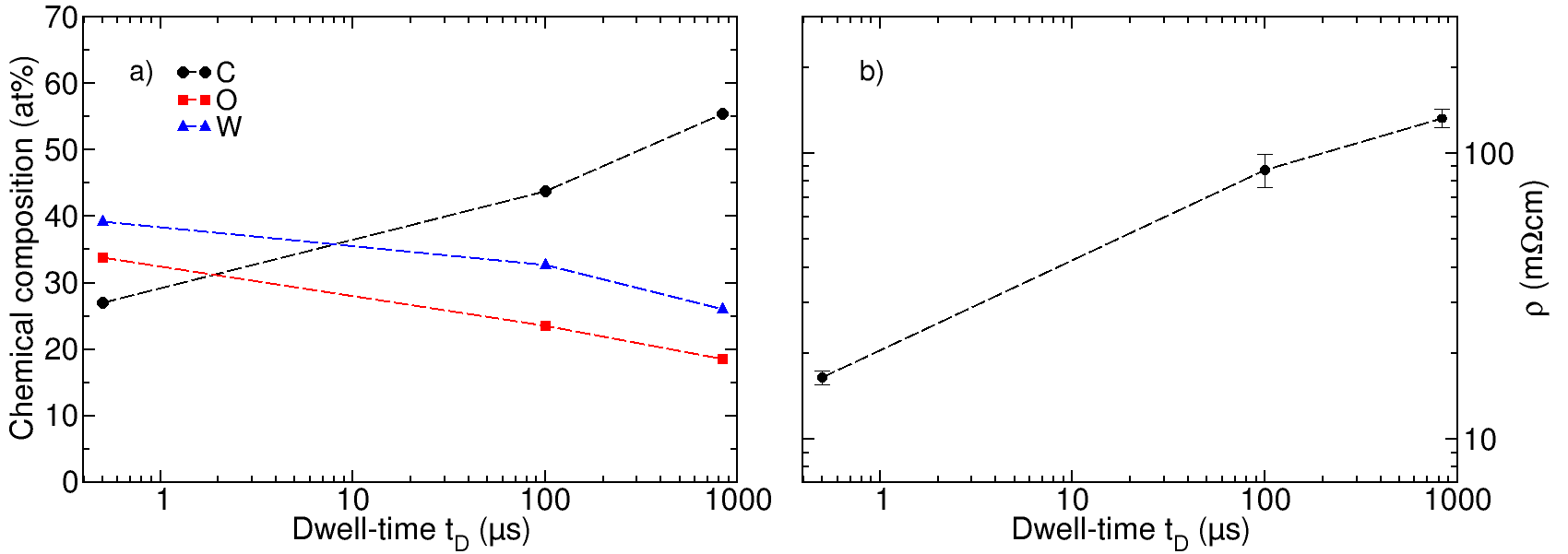}
	\caption{Chemical composition of sample 1 ($t_D = 100~\mu s$), sample 2 ($t_D = 0.5~\mu s$) and sample 3 ($t_D = 831~\mu s$). EDX measurements were performed on separate $2 \times 2~\mu m^2$ samples b) Resistivity of samples 1, 2 and 3: By solely varying the deposition parameters dwell-time and pitch as obtained from GA experiments, resistivity of W-C-O samples can be tuned by one order of magnitude.}
	\label{fig:Figure4}
\end{figure}

For the purpose of characterizing the chemical composition of the different samples energy dispersive x-ray spectroscopy (EDX) was performed. EDX measurements were carried out on $2 \times 2~\mu m^2$ reference structures deposited with the identical parameters used for the conductance measurements. In Figure \ref{fig:Figure4}a the results of the EDX measurements are displayed. Sample 2 has the highest metal content of 39.2~at\% W, whereas the metal content decreases for reference sample 1 (32.7~at\% W) and sample 3 (26.0~at\% W). Apparently a difference of more than 13~at\% between the intentionally optimal and the worst parameter set can be observed. In addition the carbon content in the deposits increases from sample 2 to sample 3, whereas the oxygen content is reduced. The corresponding resistivity of the different samples was calculated from the conductance measurements in Figure \ref{fig:Figure3}b in combination with AFM measurements of the deposits. As already indicated by the result of the EDX measurements the resistivity of the tungsten deposits is reduced by one order of magnitude for the optimized GA parameters compared to the GA parameters for lowest conductance. The results for the GA optimization for the $\rm W(CO)_6$ deposits are summarized in Table \ref{tb:Table1}.

%%%%%%%%%%%%%%%%%%%%%%%%%%%%%%%%%%%%%%%%%%%%%%%%%%%%%%%%%%%%%%%
% Figure 4
%%%%%%%%%%%%%%%%%%%%%%%%%%%%%%%%%%%%%%%%%%%%%%%%%%%%%%%%%%%%%%%
%%%%%%%%%%%%%%%%%%%%%%%%%%%%%%%%%%%%%%%%%%%%%%%%%%%%%%%%%%%%%%%

%%%%%%%%%%%%%%%%%%%%%%%%%%%%%%%%%%%%%%%%%%%%%%%%%%%%%%%%%%%%%%%
% Table 1
%%%%%%%%%%%%%%%%%%%%%%%%%%%%%%%%%%%%%%%%%%%%%%%%%%%%%%%%%%%%%%%

\begin{table}
\small
\caption{Summary of parameters used for deposition of samples 1 (reference), 2 (GA optimization for highest conductance) and 3 (GA optimization for lowest conductance). The reference sample was deposited with fixed values for dwell-time and pitch whereas the dwell-time for sample 2 was varied by the GA in the range of $t_D = 0.2-1500~ \mu s$ at fixed pitch of $p_x = p_y = 40~nm$. For sample 3, dwell-time and pitch were both allowed to vary in the range of $t_D = 0.2-1500~\mu s$ and $p_x$, $p_y = 30-200~nm$. The GA optimization was performed for 6 optimization cycles each comprising 8 parameter sets which were deposited between Cr/Au electrodes using a dose of $3~\frac{nC}{\mu m^2}$ per parameter set. The parameters obtained from the \textit{in-situ} experimental GA analysis were used to deposit another set of samples with a dose of $27~\frac{nC}{\mu m^2}$ and $A=3\times7~\mu m^2$, which were analyzed by means of AFM and electrical I(V)-measurements to obtain resisitivity of the samples. The chemical composition was determined by EDX-measurements which were performed on separate $2 \times 2~\mu m^2$ samples to prevent changing the conductivity of the samples for further electrical measurements. All other deposition parameters were kept fixed: $U=5~kV$, $I_{nominal} = 6.3~nA$} 
\label{tb:Table1} 
\begin{tabular}{|c|ccc|ccc|ccc|c|c|}\hline
\bfseries Sample &  \multicolumn{3}{c|}{\bfseries Parameters varied} & \multicolumn{3}{c|}{ \bfseries Parameters used} & \multicolumn{3}{c|}{\bfseries Chemical} & \multirow{2}{*}{\bfseries Resistivity} & \multirow{2}{*}{\bfseries Height}\\
\bfseries Nr. & \multicolumn{3}{c|}{\bfseries by GA} & \multicolumn{3}{c|}{\bfseries for deposition} & \multicolumn{3}{c|}{\bfseries composition} & & \\ \hline
\multirow{2}{*}{\#}&$t_D$ & $p_x $ & $p_y$ &$t_D$ & $p_x$ & $p_y$ & W & C & O & $\rho$ & h \\ & ($\mu s$) & (nm) & (nm) & ($\mu s$) & (nm) & (nm) & (at\%) & (at\%) & (at\%) & (m$\Omega$cm) & (nm) \\\hline
1 & - & - & - & 100 & 40 & 40 & 32.7 & 43.8 & 23.5 & 87.7 & 32 \\\hline
2 & 0.2 - 1500 & - & - & 0.5 & 40 & 40 & 39.2 & 27.0 & 33.8 & 16.5 & 36 \\\hline
3 & 0.2 - 1500 & 30 - 200 & 30 - 200 & 831 & 35 & 150 & 26.0 & 55.4 & 18.6 & 133.3 & 25 \\\hline
\end{tabular}
\end{table}
%%%%%%%%%%%%%%%%%%%%%%%%%%%%%%%%%%%%%%%%%%%%%%%%%%%%%%%%%%%%%%%

%%%%%%%%%%%%%%%%%%%%%%%%%%%%%%%%%%%%%%%%%%%%%%%%%%%%%%%%%%%%%%%
% Discussion
%%%%%%%%%%%%%%%%%%%%%%%%%%%%%%%%%%%%%%%%%%%%%%%%%%%%%%%%%%%%%%

\section{Discussion}

For the thus far presented case of $\rm W(CO)_6$, the great success of the GA optimization process is due to the fact that the metal content of the deposits can be tuned over a wide range and strongly depends on the deposition parameters which is known to be the case for many carbonyl-based precursors (e.g. $\rm W(CO)_6$ \cite{Porrati2009, Hoyle1993, Huth2009}, $\rm Co_2(CO)_8$ \cite{Serrano-Ramon2011, Utke2005} and $\rm Fe(CO)_5$ \cite{Lukasczyk2008, Shimojo2006}). With regard to the two process parameters dwell-time and pitch the FEBID process can in general be divided into two deposition regimes. For small dwell-times and larger pitches the electron induced dissociation reactions are locally limited by the number of incident electrons (reaction rate limited regime (RRL)). However, if the dwell-time is large and a small pitch is used the reactions are limited by the number of available precursor molecules (mass transport limited regime (MTL)). In most cases the electron-induced complete dissociation of a precursor molecule is not a single-step process but requires several electron-precursor interactions \cite{Dorp2008, Dorp2009}. Therefore in the RRL regime precursor molecules are not dissociated completely leading either to an implantation of non-dissociated precursor molecules or reaction by-products into the deposit but also allowing reaction by-products such as, e.g., CO groups to diffuse away from the electron impact area, desorb and finally be removed from the vacuum chamber. In the MTL regime due to the large number of locally available electrons, precursor molecules are rapidly depleted leaving enough electrons to dissociate reaction by-products which can be incorporated as non-metallic impurities into the deposit. With regard to our GA experiments RRL-like conditions \cite{Fowlkes2010} were fulfilled for sample 2 which was optimized by the GA for maximum conductance. As it is evident from the ratio of W:C:O = 1:0.69:0.86 obtained by the EDX measurements, for a short dwell-time of $0.5~\mu s$ the electron stimulated decomposition of the W-precursor and its surrounding CO ligands is very efficient as only $14.3\%$ and $11.5\%$ of oxygen and carbon atoms, respectively, of the original $W(CO)_6$ molecules are incorporated into the deposit. These findings suggest that due to the limited number of electrons available in the RRL regime the majority of volatile CO by-products can be removed during the FEBID process leading to a deposit with a high metal content. On the contrary, for a dwell-time of $831~\mu s$ the growth regime shifts to MTL regime where the replenishment rate of precursor molecules is too low leading to further electron stimulated dissociation of CO. The result is a strongly enhanced carbon content of $55.4~at\%$ in the deposit accompanied by a decrease of tungsten and oxygen to $26.0~at\%$ and $18.6~at\%$, respectively. Furthermore, the oxygen content of the deposits is coupled to the amount of tungsten indicating that tungsten-oxide is formed (Figure \ref{fig:Figure4}b). The strong increase of carbon in the deposits with decreasing oxygen content can be explained by the electron-induced decomposition of CO groups, which is in accordance with several studies on electron-induced dissociation of adsorbed and gaseous CO molecules \cite{Moore1961,Lambert1973}. Furthermore the studies show that carbon remains at the surface whereas oxygen is liberated which is in agreement with our measurements. In order to describe the observed increase of conductance it is not sufficient to only regard the metal content alone as the growth rate can also have a significant impact. However, as depicted in Table \ref{tb:Table1} AFM measurements reveal that the height of samples 1-3 varies by a factor of 1.44  corresponding to a monotonic increase of height with decreasing dwell time from 25 nm to 36 nm for samples 3 and 2, respectively. Thus, for the presented case of $W(CO)_6$ the growth rate only has a minor impact on conductance of the different samples.

\begin{figure}
	%\centering
	\sglcolfigure{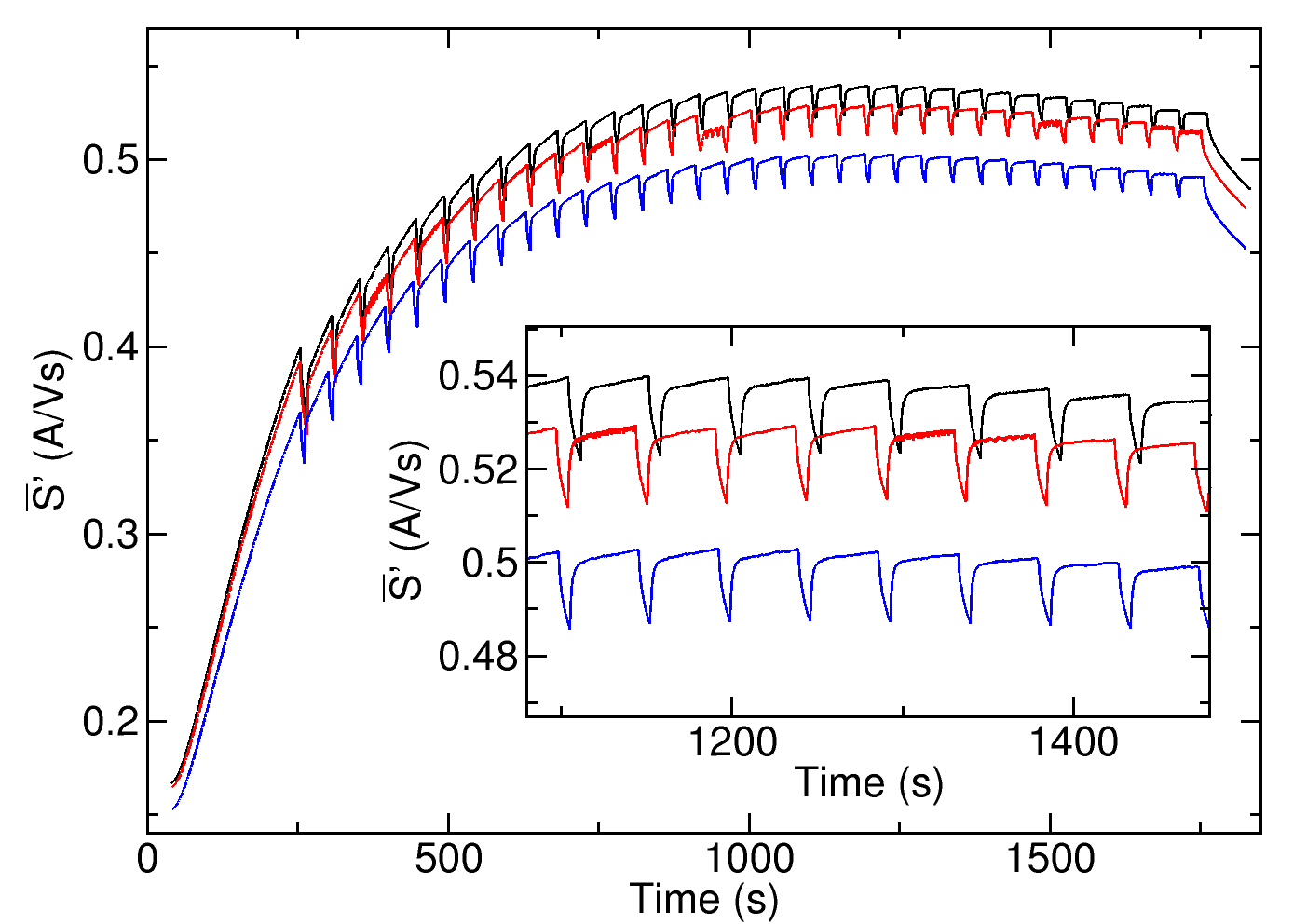}
	\caption{Time-dependent rate of change of conductance for Pt-C deposits - The GA is applied for the optimization of conductance during post-irradiation with electrons ($U = 5~kV$, $I_{nominal} = 6.3~nA$). Reference sample (blue): $t_D=1~\mu s$, $p_x=p_y=40~nm$, sample for GA dwell-time optimization (red): $t_D = 0.2-1500~\mu s$, $p_x=p_y=40~nm $, sample for GA pitch optimization (black): $t_D=1~\mu s$, $p_x,p_y=10-100~nm$. A variation of the beam-parameters dwell-time and pitch during post-growth electron treatment does not influence the rate of change of conductance during e-beam irradiation for Pt-C deposits compared to the reference (inset). The offsets in the conductance data result from small variations of conductance of the seed layer.}
	\label{fig:Figure5}
\end{figure}

The results of the GA optimization presented in this work for a precursor sensitive to the deposition parameters are extremely promising. Nevertheless, there are precursors known for the FEBID process for which the chemical composition is almost independent of the deposition parameters dwell-time and pitch. A prominent example is $\rm MeCpPt(Me)_3$. However, in this case it could be shown that the resulting Pt-C deposits are very sensitive to post-treatment either by annealing \cite{Tsukatani2005, Langford2007, Botman2009a} or electron-beam irradiation \cite{Schwalb2010a, Sachser2011, Porrati2011, Plank2011}, which can result in an increase of conductivity of many orders of magnitude. In order to investigate the influence of the GA for such a post-treatment process of FEBID deposits several Pt-C test-structures were fabricated via FEBID using identical depostion parameters ($U=5~kV$, $I=1.6~nA$, $t_D=1~\mu s$, $p_x=40~nm$, $p_y=40~nm$) and an electron dose of $30~nC/\mu m^2$. This results in a height of approximately 120~nm of the deposits, ensuring a complete penetration of the deposit by electrons. As proposed by Plank et al. \cite{Plank2011} RRL-like conditions as best initial conditions for e-beam curing were used for the deposition of Pt-C deposits, as non-dissociated precursor molecules are incorporated in the deposit. Afterwards each of the identical deposits was irradiated with the electron-beam of the SEM using: (1) standard parameters serving as a reference sample ($t_D=1~\mu s$, $p_x=p_y=40~nm$), (2) GA for dwell-time optimization ($t_D = 0.2-1500~\mu s$, $p_x=p_y=40~nm $), and (3) GA for pitch optimization $t_D=1~\mu s$, $p_x,p_y=10-100~nm$ (Figure \ref{fig:Figure5}). As can be seen in Figure \ref{fig:Figure5}, in contrast to the previous experiments for the deposition of $\rm W(CO)_6$, the variation of the irradiation parameters for dwell-time and pitch does not influence the rate of change of conductance over time which is in all cases very strong. Therefore, for electron post-treatment of samples deposited with the Pt-based precursor $\rm MeCpPt(Me)_3$ no parameter sets resulting in a significantly faster enhancement of the conductance could be identified with the GA. 

%%%%%%%%%%%%%%%%%%%%%%%%%%%%%%%%%%%%%%%%%%%%%%%%%%%%%%%%%%%%%%%
% Figure 5
%%%%%%%%%%%%%%%%%%%%%%%%%%%%%%%%%%%%%%%%%%%%%%%%%%%%%%%%%%%%%%%

According to Plank et al. the post-growth irradiation-induced dissociation of incorporated molecules leads to the creation of small Pt-crystallites between existing Pt-crystals in the nanogranular structure of Pt-C or to a growth of the previously present Pt crystallites leading to a reduction of the intergrain distance and therefore to decreasing resistivity \cite{Plank2011}. We found that, as already shown in previous experiments \cite{Porrati2011}, the resistivity could be reduced during e-beam curing, however, independent of dwell-time and pitch. This can be expected because precursor depletion as the dominant factor during deposition does not play a role during e-beam curing. Furthermore, effects like the growth of existing Pt crystals should depend on the electron dose rather than on parameters such as dwell-time or pitch for post-irradiation of samples at fixed dose.

%%%%%%%%%%%%%%%%%%%%%%%%%%%%%%%%%%%%%%%%%%%%%%%%%%%%%%%%%%%%%%%

%%%%%%%%%%%%%%%%%%%%%%%%%%%%%%%%%%%%%%%%%%%%%%%%%%%%%%%%%%%%%%
% Summary
%%%%%%%%%%%%%%%%%%%%%%%%%%%%%%%%%%%%%%%%%%%%%%%%%%%%%%%%%%%%%%

\section{Conclusion}

In this work we presented the application of an evolutionary GA for the \textit{in-situ} optimization of FEBID nanostructures with regard to their electrical conductivity. By using the gradient of the measured \textit{in-situ} rate of change of conductance as a fitness parameter the GA was able to tune the metal content of tungsten deposits created from $\rm W(CO)_6$ over a large range by either targeting the highest or lowest conductance, respectively. This resulted in a difference in conductivity of one order of magnitude. This experiment highlights the effectiveness of the procedure for precursors for which the chemical composition of the deposit is sensitive to the deposition parameters. In a second experiment the GA was applied for post-treatment of Pt-C deposits obtained from the precursor $\rm MeCpPt(Me)_3$ by electron-beam irradiation. For this system the GA revealed that solely the applied electron dose and not specific irradiation parameters leads to the observed strong increase of conductance over time.

The presented technique can be applied to all beam-induced deposition processes and has great potential for further optimization or tuning of parameters for nanostrucures prepared by FEBID or related techniques. In particular finding optimized deposition parameters for new precursor materials, which in general is a very time-consuming and often an arbitrary process, can be achieved in a fast and efficient way. The GA's independence of the mechanism responsible for the enhancement of conductance (e.g. increase of metal content, changes of height of the deposit, structural or phase changes, etc.) and its adaption to every experimental circumstance with direct feedback promises significant potential for future FEBID research. Furthermore, the application of the GA is not restricted to the optimization of conductance but can also be applied to e.g. optimize dielectric properties of FEBID deposits using capacative measurements or optical reflectivity. Especially it will play a major role for the analysis and optimization of FEBID binary systems that have been recently adressed \cite{Winhold2011, Porrati2012, Che2005}. Some of us were able to stabilize an amorphous, metastable $\rm Pt_2Si_3$ phase showing a maximum of conductivity compared to other Pt-Si samples with different stoichiometric proportions of platinum and silicon \cite{Winhold2011}. In follow-up experiments it will be shown, that the GA can be applied to obtain deposition parameters for binary systems e.g. Pt-Si or Co-Pt, which automatically lead to the formation of binary phases with highest conductivity.

%%%%%%%%%%%%%%%%%%%%%%%%%%%%%%%%%%%%%%%%%%%%%%%%%%%%%%%%%%%%%
% Acknowledgements
%%%%%%%%%%%%%%%%%%%%%%%%%%%%%%%%%%%%%%%%%%%%%%%%%%%%%%%%%%%%%
\begin{acknowledgements}
The authors acknowledge financial support by the Beilstein-Institut, Frankfurt/Main, Germany within the research collaboration NanoBiC.
\end{acknowledgements}

%%%%%%%%%%%%%%%%%%%%%%%%%%%%%%%%%%%%%%%%%%%%%%%%%%%%%%%%%%%%%
%\section{Supplementary}
%%%%%%%%%%%%%%%%%%%%%%%%%%%%%%%%%%%%%%%%%%%%%%%%%%%%%%%%%%%%%

%The pause allowed the program to read in the next parameter set as a streamfile with other process parameters. 

%%%%%%%%%%%%%%%%%%%%%%%%%%%%%%%%%%%%%%%%%%%%%%%%%%%%%%%%%%%%%
%\section{References}
%%%%%%%%%%%%%%%%%%%%%%%%%%%%%%%%%%%%%%%%%%%%%%%%%%%%%%%%%%%%%

\bibliography{bib_beilstein_journal_of_nanotechnology2013}
\end{document}